\documentclass[prd,aps,twocolumn]{revtex4}
\usepackage{amsmath}
\usepackage{graphicx}
\usepackage{amssymb}
\usepackage{hyperref}
\usepackage{color}

\definecolor{verde}{rgb}{0,0.5,0}
\def\blu{\color{blue}}

\def\be{\begin{equation}}
\def\ee{\end{equation}}
\def\bea{\begin{eqnarray}}
\def\eea{\end{eqnarray}}
\def\be{\begin{equation}}
\def\ee{\end{equation}}
\def\ba{\begin{align}}
\def\ea{\end{align}}

\def\noi{\noindent}

\newcommand\lsim{\mathrel{\rlap{\lower4pt\hbox{\hskip0.5pt$\sim$}}
    \raise1pt\hbox{$<$}}}
\newcommand\gsim{\mathrel{\rlap{\lower4pt\hbox{\hskip0.5pt$\sim$}}
    \raise1pt\hbox{$>$}}}

\begin{document}

%
%
%
\renewcommand{\topfraction}{0.99}
\renewcommand{\bottomfraction}{0.99}

 \title{Searching for Fossil Fields in the Gravity Sector}
\author{Emanuela Dimastrogiovanni$^a$, Matteo Fasiello$^{b}$, Gianmassimo Tasinato$^c$}

\affiliation{$^a$School of Physics, The University of New South Wales, Sydney NSW 2052, Australia}
\affiliation{$^b$Institute of Cosmology and Gravitation, University of Portsmouth, Portsmouth, PO1 3FX, U.K.}
\affiliation{$^c$Department of Physics, Swansea University, Swansea, SA2 8PP, U.K.}

\begin{abstract}
\noi Evidence for the presence of extra fields during inflation may be found in the anisotropies of the scalar and tensor spectra across a vast range of scales. Indeed, beyond the single-field slow-roll paradigm, a long tensor mode can modulate the power spectrum inducing a sizable  quadrupolar anisotropy.  We investigate how  this dynamics plays out for the tensor two-point correlator. The resulting quadrupole stores information on squeezed tensor non-Gaussianities, specifically those  sourced by the extra field content and responsible for the breaking of so-called consistency relations. We underscore the potential of anisotropies as a probe of new physics: testable at CMB scales through the detection of B-modes, they are accessible at smaller scales via interferometers and pulsar timing arrays.

\end{abstract}

\maketitle
\noindent
\noindent

\section {Introduction}
\label{intro}

\begin{figure}[h!]
	\begin{center}
		\includegraphics[width=7.cm]{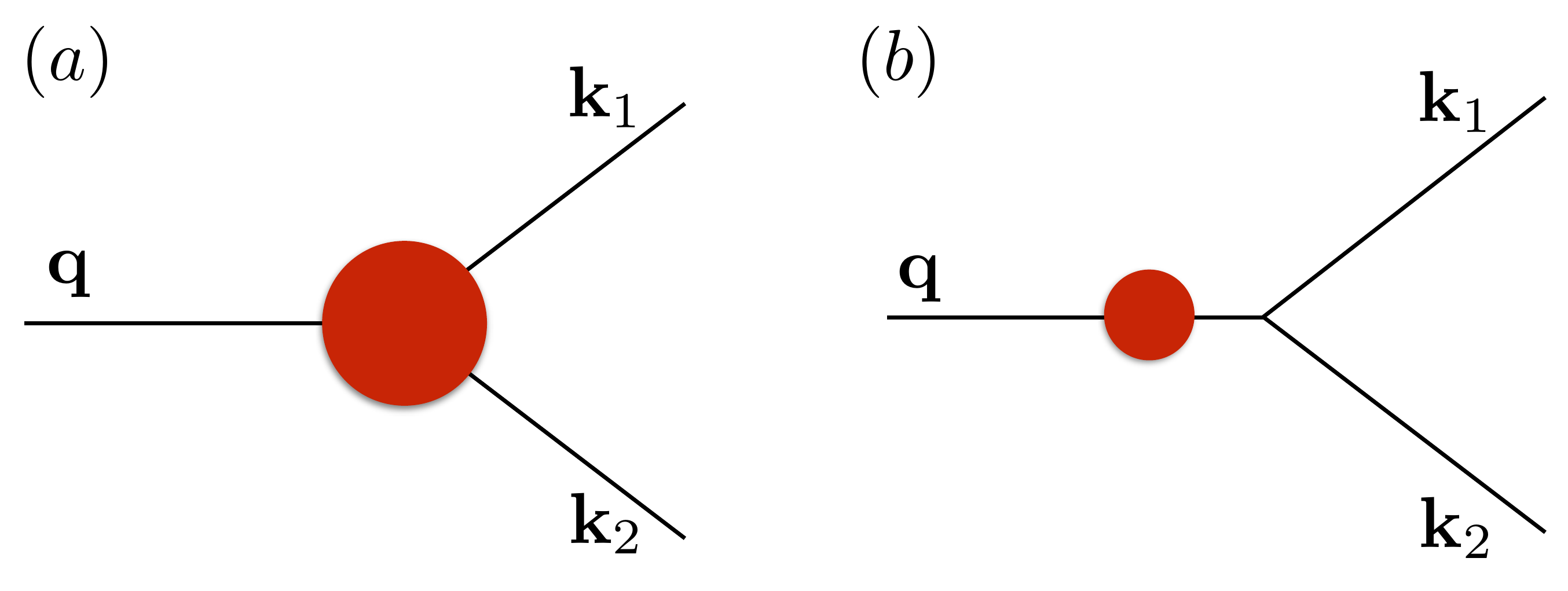}
		\caption
		{\it Diagrammatic representation of a tensor three-point function mediated by extra $\sigma$ fields. Red circles indicate interactions between the tensor modes of the metric and $\sigma$ fields. Diagrams leading to a non-trivial consistency-relation-breaking bispectrum are the subset of those in (a) that cannot be simplified to the form in (b). Indeed, the latter diagram is sensitive to information on $\sigma$, which is already probed via the standard tensor two-point function.				}
		\label{fig2}
	\end{center}
\end{figure}

\noi The detection of gravitational waves from \textcolor{black}{black hole mergers and from colliding neutron stars \cite{Abbott:2016blz}} has ushered in a new era for astronomy. The same is bound to happen for early universe cosmology upon observing (evidence of) a primordial tensor signal.
In particular, probes of the gravity sector hold a great discovery potential when it comes to inflationary physics. Detection of CMB B-modes polarisation would, in standard single-field slow-roll scenarios, precisely identify the energy scale of inflation. Crucially, gravitational probes can access precious information on the early acceleration phase also in the case of multi-field inflation. \\
A non-minimal inflationary field content is not only possible, but perhaps even likely \cite{Baumann:2014nda}. String theory realisations of the acceleration mechanism typically result in  extra dynamics due, for example, to compactifications moduli.  Axion particles as well as  Kaluza-Klein modes and gauge fields can also be accommodated. The extra content acts as a source of the standard inflationary scalar and tensor fluctuations. An interesting phenomenology ensues whereby tensor fluctuations, sometimes sourced already at linear order,  may deliver a non-standard 
 tensor power spectrum exhibiting a marked scale dependence, features, and in specific cases \cite{Anber:2009ua,Barnaby:2010vf,Adshead:2012kp,Dimastrogiovanni:2016fuu,Pajer:2013fsa,Garcia-Bellido:2016dkw} a chiral signal. {Additionally, a similar dynamics is arrived at by employing new (broken) symmetry patterns \cite{Endlich:2012pz} or so-called non-attractor phases for the inflationary mechanism 
 \cite{Ozsoy:2019slf,Mylova:2018yap}.}\\
Perhaps the most sensitive probe of extra physics is the  (scalar/tensor/mixed) bispectrum. Its amplitude and momentum dependence can be mapped onto specific properties of the inflationary Lagrangian. Remarkably, the soft momentum limit of the bispectrum contains detailed information \cite{Arkani-Hamed:2015bza} on the mass, the spin, and (implicitly) the couplings of extra fields. The existence of a non-trivial squeezed  bispectrum contribution, mediated by the extra content, may be inferred already at the level of the tensor power spectrum. Indeed, in what we shall call the \textit{ultra-squeezed} configuration, a  long tensor mode induces a position-dependence in the short \footnote{Clearly, the bispectrum is at the origin of this effect: it is its momentum conservation rule that forces the two modes correlated with the long tensor to be short. } modes power spectrum. {In this context, a non-trivial bispectrum corresponds to one that modifies so-called consistency relations (CRs).   These are maps between ``soft" limits of N+1-point functions and their lower order  counterpart that result from a residual diffeomorphism in the description of the physical system. Standard inflationary CRs are modified in the presence of e.g. non-Bunch-Davies initial conditions, independent modes that transform non-linearly under the diffeomorphism, alternative symmetry breaking patterns,   etc \cite{Hinterbichler:2013dpa}. A prototypical example of modified CRs stems from the presence of extra fields during inflation. Interactions mediated by (see Fig.~\ref{fig2}) the extra  $\sigma$ content are precisely those that can modify CRs. As a result of CRs breaking, the $\sigma$-mediated leading contribution to the squeezed bispectrum is physical, i.e. it cannot be gauged away.  
}

\textcolor{black}{In what follows, we analyse the power spectrum of tensor fluctuations in the presence of a long-wavelenth tensor mode as a probe of squeezed tensor non-Gaussianity \footnote{See e.g. Sec.~5 of \cite{Bartolo:2018qqn} for a review on tensor non-Gaussianity from inflation}. The modulation effect can occur at widely different scales, from the CMB all the way to regimes accessible via interferometers. In Sec.~\ref{sec2} we analyse in detail how the squeezed tensor
bispectrum induces a quadrupolar anisotropy in the tensor power spectrum. In Sec.~\ref{sec:3}, building on the recent important  works \cite{Bartolo:2018evs,Bartolo:2018rku}, we elaborate on the fact that such observable is not plagued by the suppression effects that prevent a direct measurement of tensor non-Gaussianity at interferometer scales. We conclude with
Sec.~\ref{conclusions}.
}

\section{Quadrupolar anisotropy}
\label{sec2}

\noindent  If inflation predicts a non-trivial \footnote{We refer the reader interested in explicit models to the work in \cite{Bordin:2018pca,Dimastrogiovanni:2018gkl,Dimastrogiovanni:2015pla,Emami:2015uva} and references therein, where intriguing bispectrum signatures emerge from contributions mediated by additional degrees of freedom. See also Fig.~\ref{fig2}a for a diagrammatic representation.} tensor-scalar-scalar bispectrum in the squeezed limit (the tensor being the soft mode), a quadrupolar anisotropy is induced by the long-wavelength tensor mode in the observed \textsl{local} scalar power spectrum \cite{Dai:2013kra}.
 In close analogy to the procedure developed in \cite{Dai:2013kra} for the scalar case, we now show (see also \cite{Ricciardone:2017kre,Dimastrogiovanni:2018uqy}) that the tensor power spectrum is modulated by a long-wavelength tensor fluctuation. The quadrupolar anisotropy of the tensor power spectrum is then an observable sensitive to the squeezed limit of tensor non-Gaussianity.
\\

\noindent The starting point is the correlation between two tensor modes in the presence of a long-wavelength mode. We express the primordial spin-2 tensor fluctuation around a conformally flat, FRW metric in terms of its Fourier modes as
\begin{equation}
\hat{\gamma}_{ij}(\textbf{x},\tau)=\int\frac{d^{3}k}{(2\pi)^{3}}e^{i\textbf{k}\cdot\textbf{x}}\hat{\gamma}_{\textbf{k},ij}(\tau)\,,
\end{equation}
adopting  the mode decomposition
\begin{align}
\hat{\gamma}_{\textbf{k},ij}(\tau)&=\sum_{\lambda=R,L}\epsilon_{ij}^{\lambda}(\hat{\textbf{k}})\hat{\gamma}_{\textbf{k}}^{\lambda}(\tau)\,,\\
 \hat{\gamma}_{\textbf{k}}^{\lambda}(\tau)&=a_{\textbf{k}}^{\lambda}\gamma^{\lambda}_{k}(\tau)+a_{-\textbf{k}}^{\lambda\,\dagger}\gamma^{\lambda\,*}_{k}(\tau)\,.
\end{align}
The creation and annihilation operators $a_{\textbf{k}}^{\lambda\,\dagger},\,a_{\textbf{k}}^{\lambda}\,$, satisfy the standard commutation
relations. The polarisation tensors $\epsilon_{ij}^{\lambda}(\hat{\textbf{k}})$ are transverse and traceless, are normalised such
that $\epsilon_{ij}^{R}(\hat{\textbf{k}})\epsilon_{ij}^{R}(-\hat{\textbf{k}})=\epsilon_{ij}^{L}(\hat{\textbf{k}})\epsilon_{ij}^{L}(-\hat{\textbf{k}})=1$,
and moreover
$\epsilon_{\ell m}^{R}(-\hat{\textbf{k}})=\epsilon_{\ell m}^{R\,*}(\hat{\textbf{k}})=\epsilon_{\ell m}^{L}(\hat{\textbf{k}})=\epsilon_{\ell m}^{L\,*}(-\hat{\textbf{k}})$.  In the absence of significant modulations induced  by couplings with long-wavelength modes, the tensor power spectrum and bispectrum  --
for models that do not violate isotropy nor parity symmetry -- 
are given
by
\bea
\langle  \hat{\gamma}_{\textbf{k}_{1}}^{\lambda_{1}}\hat{\gamma}_{\textbf{k}_{2}}^{\lambda_{2}}\rangle
\,
&\equiv&\, (2\pi)^{3}\,\delta^{\lambda_{1}\lambda_{2}}\,\delta^{(3)}(\textbf{k}_{1}+\textbf{k}_{2})P^{\lambda_{1}}_{\gamma}(k_{1})\,,
\\
\langle \hat{\gamma}^{\lambda_{3}}_{\textbf{q}} \hat{\gamma}_{\textbf{k}_{1}}^{\lambda_{1}}\hat{\gamma}_{\textbf{k}_{2}}^{\lambda_{2}} \rangle 
&\equiv& (2\pi)^{3}\delta^{(3)}(\textbf{k}_{1}+\textbf{k}_{2}+\textbf{q})B_{\gamma}^{\lambda_{1}\lambda_{2}\lambda_{3}}(\textbf{k}_{1},\textbf{k}_{2},\textbf{q})\,.
\nonumber \\
\eea
The general expression that accounts for modulation due to the coupling with long-wavelength tensor modes is instead
 \cite{Jeong:2012df}:
\begin{eqnarray}\label{q1}
&&\langle  \hat{\gamma}_{\textbf{k}_{1}}^{\lambda_{1}}\hat{\gamma}_{\textbf{k}_{2}}^{\lambda_{2}}\rangle_{\gamma_{L}}
\,
\equiv\, (2\pi)^{3}\,\delta^{\lambda_{1}\lambda_{2}}\,\delta^{(3)}(\textbf{k}_{1}+\textbf{k}_{2})P^{\lambda_{1}}_{\gamma}(k_{1})+\nonumber\\
&& \sum_{\lambda_{3}}\int_{|\vec q|<q_L}d^{3}q\,\delta^{(3)}(\textbf{k}_{1}+\textbf{k}_{2}+\textbf{q}) \gamma^{*\lambda_{3}}_{\textbf{q}}\,\frac{B_{\gamma}^{\lambda_1 \lambda_2 \lambda_3}(\textbf{k}_{1},\textbf{k}_{2},\textbf{q})}{P^{\lambda_{3}}_{\gamma}(q)}\,,
\nonumber\\
\end{eqnarray}
where $q_L$ is a cut-off on whose size we shall soon elaborate. It suffices here to say that it is ensuring we integrate only over the squeezed configurations of the bispectrum $B_{\gamma}$. In standard single-field inflationary models,  the leading terms in  $B_{\gamma}$ are related to the scale dependence of the (short) tensor power spectrum through consistency relations \cite{Maldacena:2002vr}. As a result, their effect can be removed by an appropriate gauge transformation  (see e.g. \cite{Gerstenlauer:2011ti,Giddings:2011zd,Dai:2013kra,Dai:2015rda,Sreenath:2014nka}). In models where consistency relations are modified or broken, $B_{\gamma}$ accesses directly new physical information stored in the squeezed tensor bispectrum. It is in the case of the latter set of models and, in particular, of their consistency-relation-breaking contributions to the bispectrum, that our analysis becomes especially relevant. It will be convenient in what follows to use the quantity $\tilde B$, defined as 
\begin{align}\label{q4}
& B_{\gamma}^{\lambda_{1}\lambda_{2}\lambda_{3}}|_{ q \ll k_{1,2}} \simeq- \delta^{\lambda_{1}\lambda_{2}}\epsilon_{\ell m}^{\lambda_{3}}(\hat{q})\hat{k}_{1\ell}\hat{k}_{2m}\,\tilde{B}(\textbf{k}_{1},\textbf{k}_{2},\textbf{q})\,,
\end{align}
in order to make the dependence on polarisation indices  explicit. The quantity in $\tilde{B}$ can be parametrically large and lead to observable effects. To make the physical consequences of such modulation more manifest, it is useful to express the tensor two point function as
\begin{align}
&\langle \hat{\gamma}_{ij}(\textbf{x}_{1})\hat{\gamma}_{ij}(\textbf{x}_{2}) \rangle_{\gamma_{L}}=\nonumber\\ 
&\Big\langle \sum_{\lambda_{1},\lambda_{2}}\int\frac{d^{3}k_{1}}{(2\pi)^{3}}e^{i\textbf{k}_{1}\cdot\textbf{x}_{1}}\epsilon_{ij}^{\lambda_{1}}(\hat{\textbf{k}}_{1})\hat{\gamma}_{\textbf{k}_{1}}^{\lambda_{1}}   \int\frac{d^{3}k_{2}}{(2\pi)^{3}}e^{i\textbf{k}_{2}\cdot\textbf{x}_{2}}\epsilon_{ij}^{\lambda_{2}}(\hat{\textbf{k}}_{2})\hat{\gamma}_{\textbf{k}_{2}}^{\lambda_{2}}\Big\rangle_{\gamma_{L}}\nonumber\\
&=\int\frac{d^{3}k_{1}}{(2\pi)^{3}}\,e^{i\textbf{k}_{1}(\textbf{x}_{1}-\textbf{x}_{2})}\,P_{\gamma}(k_{1})+  {\cal S}\,,
\end{align}
where $P_{\gamma}=P_{\gamma}^{R}+P_{\gamma}^{L}$. Upon introducing the new coordinates 
\bea
\textbf{k}\equiv  \frac{\textbf{k}_{2}-\textbf{k}_{1}}{2}\,,\quad \textbf{p}\equiv \textbf{k}_{1}+\textbf{k}_{2}, \nonumber\\  \textbf{x}_{c}\equiv  \frac{\textbf{x}_{1}+\textbf{x}_{2}}{2}\,,\quad \textbf{x}\equiv \textbf{x}_{2}-\textbf{x}_{1},
\eea
and using $\delta^{(3)}(\textbf{k}_{1}+\textbf{k}_{2}+\textbf{q})=\delta^{(3)}(\textbf{p}+\textbf{q})$, one finds
\begin{align}
\label{q2}
{\cal S}=&\sum_{\lambda_{1}\lambda_{2}\lambda_{3}}\int\frac{d^{3}k}{(2\pi)^3}\frac{d^{3}p}{(2\pi)^3} \times \nonumber\\&\,\,\epsilon_{ij}^{\lambda_{1}}\left(\widehat{\frac{\textbf{p}}{2}-\textbf{k}}\right)\epsilon_{ij}^{\lambda_{2}}\left(\widehat{\frac{\textbf{p}}{2}+\textbf{k}}\right)\,e^{i\left(\textbf{p}\cdot\textbf{x}_{c}+\textbf{k}\cdot\textbf{x}_{}\right)}P_{\gamma}^{}(k)\nonumber\\
&\times\gamma^{\lambda_{3}\,*}_{-\textbf{p}}\frac{\tilde{B}(\textbf{k}, \,\textbf{p})}{P_{\gamma}(k)P_{\gamma}^{\lambda_{3}}(p)}\epsilon^{\lambda_{3}}_{\ell m}(-\hat{p})\hat{k}_{\ell}\hat{k}_{m}\delta^{\lambda_{1}\lambda_{2}}\,\nonumber\\
%
 =&\int\frac{d^{3}k}{(2\pi)^3} e^{i \textbf{k}\cdot\textbf{x}_{}}P_{\gamma}^{}(k)\,\mathcal{Q}_{\ell m}(\textbf{x}_{c},\,\textbf{k})\hat{k}_{\ell}\hat{k}_{m}\,,
\end{align}
where the anisotropy parameter $\mathcal{Q}_{\ell m}$ is given by
\begin{align}\label{q11}
&\mathcal{Q}_{\ell m}(\textbf{x}_{c},\textbf{k})\equiv 
\int \frac{d^{3}q}{(2\pi)^3}\, e^{i \textbf{q}\cdot\textbf{x}_{c}} \times\nonumber \\ &\sum_{\lambda_{3}}
\left[\frac{\tilde{B}(\textbf{k},\,\textbf{q})}{(1/2)P_{\gamma}(k)P_{\gamma}^{\lambda_{3}}(q)}\right]\epsilon^{\lambda_{3}}_{\ell m}(-\hat{q})\,\gamma^{*\lambda_{3}}_{-\textbf{q}}\nonumber\\=&
\int \frac{d^{3}q}{2 \pi^3}\, e^{i \textbf{q}\cdot\textbf{x}_{c}}\,f_{\rm nl}(\textbf{q},\textbf{k})\sum_{\lambda_{3}}
\epsilon^{\lambda_{3}}_{\ell m}(-\hat{q})\,\gamma^{*\lambda_{3}}_{-\textbf{q}} \; .
\end{align}
Note that in Eq.~(\ref{q11}) the following parameterisation $\tilde{B}(\textbf{k},\,\textbf{q})\,=\, f_{\rm nl}(\textbf{q},\textbf{k})\,P_{\gamma}(k)P_{\gamma}(q)$ has been adopted and the usual properties of the polarisation tensors have been used. The quantity $ f_{\rm nl}$ parameterises  the amplitude and momentum dependence of the squeezed limit of the 
tensor bispectrum. Going back to Fourier space and expressing $\textbf{x}_{1}$ and $\textbf{x}_{2}$ in terms of $\textbf{x}$ and $\textbf{x}_{c}$, one finds the following expression for the tensor power spectrum in the presence of a long-wavelength tensor mode $\gamma_{\textbf{q}}$, evaluated locally, i.e. within a volume whose linear dimension is smaller than the wavelength of the tensor ($|\textbf{x}|\ll 1/q$):
\begin{align}\label{q13}
&P_{\gamma}^{}(\textbf{k}',\textbf{x}_{c})|_{\gamma_{L}}\equiv\int d^{3}x\, e^{-i\textbf{k}'\cdot\textbf{x}}\Big\langle \hat{\gamma}_{ij}\left(\textbf{x}_{c}-\frac{\textbf{x}}{2}\right)\hat{\gamma}_{ij}\left(\textbf{x}_{c}+\frac{\textbf{x}}{2}\right) \Big\rangle_{\gamma_{L}}\nonumber\\
&= P_{\gamma}(k')\Big(1+\mathcal{Q}_{\ell m}(\textbf{x}_{c},\textbf{k}')\hat{k}'_{\ell}\hat{k}'_{m}\Big)\,.
\end{align}

\begin{figure}[h!]
\begin{center}
  \includegraphics[width=7cm]{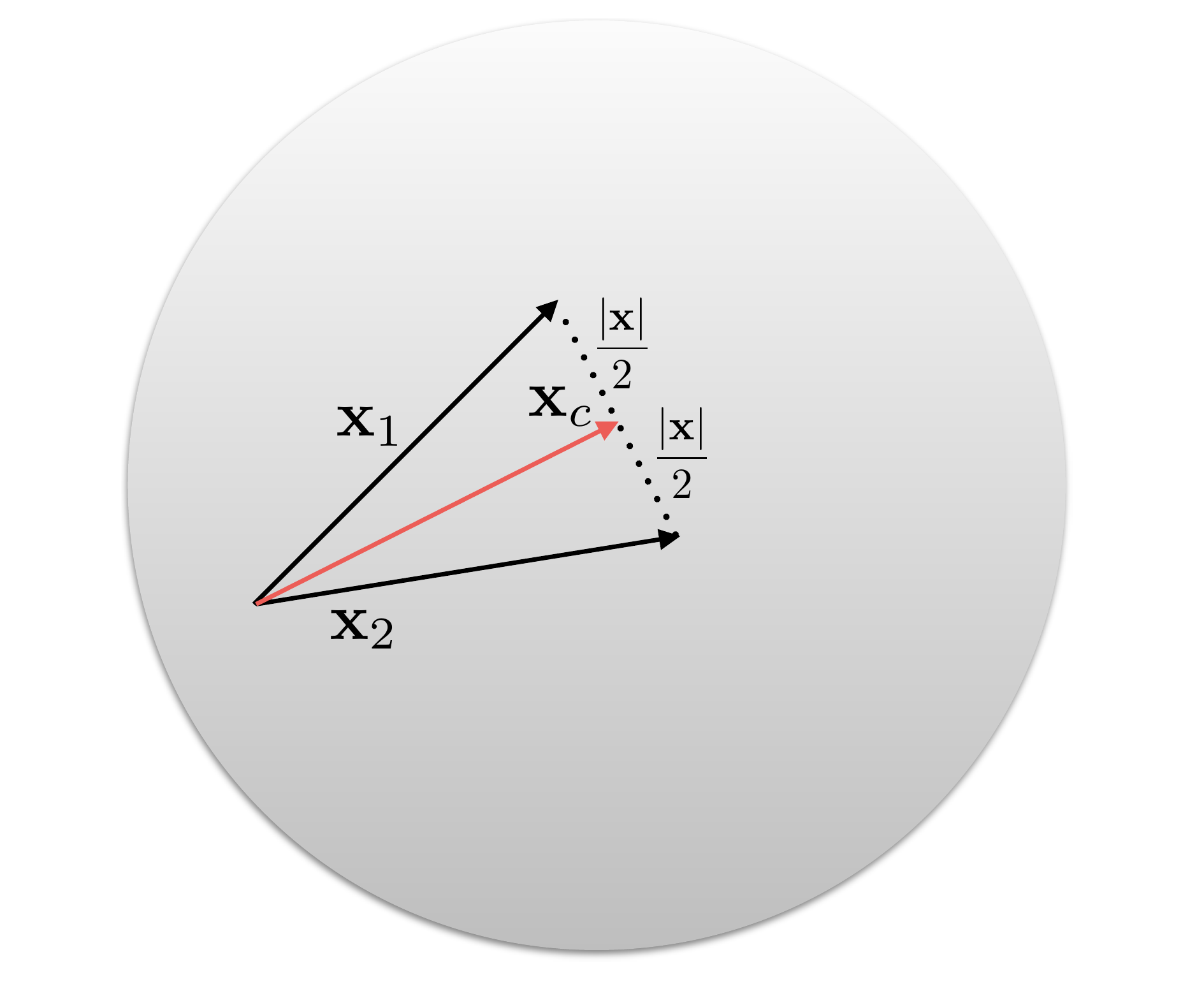}
  \caption
 {\it in Eq.~(\ref{q13}), for a given point in space $\textbf{x}_{c}$, the volume of integration is defined by a length scale much smaller than the typical variation scale of the modulating tensor mode. If inflation supports a non-zero tensor bispectrum in the squeezed configuration, the power spectrum defined locally, i.e. in the vicinity of $\textbf{x}_{c}$, has a quadrupole modulation caused by long-wavelength tensor fluctuations.}
 \label{fig1}
\end{center}
\end{figure}
\noindent Expanding the quadrupole modulation in spherical harmonics and computing its variance, one arrives at 
\begin{align}
\label{variance}
\overline{\mathcal{Q}^2}\equiv \Big\langle \sum_{m=-2}^{+2} |\mathcal{Q}_{2 m}|^2 \Big\rangle= \frac{8\pi}{15} \Big\langle  \mathcal{Q}_{ij}\mathcal{Q}^{*}_{ij}\Big\rangle\,,
\end{align}
and 
\begin{align} \label{expQij}
\Big\langle  \mathcal{Q}_{ij}\mathcal{Q}^{*}_{ij}\Big\rangle=16 \int \frac{d^2  \hat{q}}{4 \pi}\,\int_{q^{\text{min}}}^{q^{\text{max}}}\frac{dq}{q}f_{\text{nl}}^{2}( \textbf{q}, \textbf{k})\mathcal{P}_{\gamma}(q)\,,
\end{align}
where the definition  $\mathcal{P}_{\gamma}(q)\equiv q^{3}P_{\gamma}(q)/2\pi^{2}$ has been used. We pause here to comment on how the extrema of integration over $q$ are chosen. The lower value $q^{\text{min}}$ is selected as the wavenumber corresponding to the longest wavelengths that ever exited the horizon. The value of $q^{\text{max}}$ depends instead on the specific probe.  
For CMB observations, for example, $q^{\text{max}}$ is given by the smallest wavenumber probed by a given experiment. The case of interferometers will be the subject of a more detailed discussion in what follows. \\
\noi Whenever $f_{\rm nl}$ 
 and the tensor power spectrum ${\cal P}_\gamma$ are scale-invariant, Eq.~\eqref{expQij} simplifies to
\begin{align}\label{qmM}
\Big\langle  \mathcal{Q}_{ij}\mathcal{Q}^{*}_{ij}\Big\rangle=16 f_{\text{nl}}^{2}\,\mathcal{P}_{\gamma}\ln\left(\frac{q^{\text{max}}}{q^{\text{min}}}\right)
\,.
\end{align}
Models of (super)solid  \cite{Ricciardone:2017kre,Ricciardone:2016lym} and non-attractor  \cite{Ozsoy:2019slf} inflation do indeed support in some regimes an (almost) scale invariant profile for both the power spectrum and $f_{\rm nl}$. Let us focus on  Eq.~(\ref{qmM}) in the case of CMB polarisation. Recalling the definition of the tensor-to-scalar ratio as  $r\equiv\mathcal{P}_{\gamma}/\mathcal{P}_{\zeta}$,  one finds
\begin{align}
\sqrt{\overline{\mathcal{Q}^2}}=2.4\cdot 10^{-4}f_{\text{nl}}\, \sqrt{r\,\Delta N}\,,
\end{align}
where the number of e-folds between the exit of the longest mode $q^{\text{min}}$ and the exit of $q^{\text{max}}$ is $\Delta N\equiv \ln(q^{\text{max}}/q^{\text{min}})$. As an example, with $\Delta N=\mathcal{O}(1)$, a value for $\sqrt{\overline{\mathcal{Q}^2}}$ of order $0.1$  requires $f_{\text{nl}}\sqrt{r}\simeq 500$. \textcolor{black}{Given current constraints on the tensor-to-scalar ratio, this demands $f_{\text{nl}}\gtrsim 2\cdot 10^{3}$.} Observational bounds on tensor non-Gaussianity have been obtained from temperature and E-mode polarization data for a class of models predicting  \textcolor{black}{bispectra that peak in the equilateral configuration \cite{Akrami:2019izv}}: $f_{\text{nl}}^{\text{Planck}}=500\pm 1100\, (68\%\,\text{CL})$, where $f_{\text{nl}}^{\text{Planck}}\equiv B(k,k,k)/[(18/5)P_{\zeta}^{2}(k)]$. These constraints do not immediately apply to our case. However, even if enforced on our set-up, they would be  compatible with $f_{\text{nl}}\sqrt{r}=\mathcal{O}(500)$. It would be interesting to forecast the bounds that future CMB polarization experiments will be able to place on squeezed non-Gaussianity by probing the tensor quadrupolar anisotropy. We leave this to future work.

\noi If the tensor power spectrum and/or $f_{\rm nl}$ are not scale invariant, the expression for the quadrupole anisotropy is modified w.r.t. Eq.~(\ref{qmM}). As a concrete example, let us consider a tensor bispectrum mediated by a massive spin-2 field with a small speed of sound \cite{Bordin:2018pca}. In this set-up, the power spectrum of gravitational waves receives the standard vacuum contribution and the one generated by the extra field: $P_{\gamma}(k)=[(4H^{2})/(M_{\text{Pl}}^{2}k^{3})](1+\alpha_{\gamma})$. Considering  contributions to the bispectrum mediated by the extra particle, for $\alpha_{\gamma}=\mathcal{O}(1)$ one finds \cite{Dimastrogiovanni:2018gkl}:  
\begin{align}
\label{spin2}
f_{\text{nl}}\simeq 4\cdot 2^{\nu}\cdot\frac{M_{\text{Pl}}}{H}\cdot\frac{\mu}{H}\cdot\frac{\rho^{3}}{H^{3}}s^{\text{sq}}(\nu,c_{\sigma})\left(\frac{q}{k}\right)^{\frac{3}{2}-\nu}\,,
\end{align}
where $\nu\equiv\sqrt{9/4-m^{2}_{\sigma}/H^{2}}$, with $m_{\sigma}$ the mass of the spin-2 field, and $c_{\sigma}$ its sound speed. The parameters $\mu$ and $\rho$ quantify respectively the magnitude of the cubic self-interaction of the spin-2 field and of the quadratic mixing of the same field with metric tensor modes. In this set-up, for $m_{\sigma}$ of order Hubble, the square root of the quadrupole variance is typically of order $10^{-1}$ on CMB scales. This in spite of the $(q/k)^{3/2-\nu}$ suppression w.r.t. the scale-invariant case. 
We note here that models with excited initial states (see e.g. \cite{Brahma:2013rua}) may also be of particular interest for quadrupolar anisotropies. In some of these constructions $f_{\rm nl}$ scales with negative powers of $q/k$, which may easily lead to a sizable quadrupole.\\

\noindent As the above examples illustrate, the  quadrupolar asymmetry corresponding to a scale-dependent spectrum and $f_{\rm nl}$ is model-dependent. If the primordial gravitational wave (GW) spectrum has a sufficiently large amplitude at small scales (see e.g. \cite{Bartolo:2016ami} for a review of several such scenarios), primordial non-Gaussianity can act as a source for  anisotropies of the stochastic GW background. These can be detectable (see Sec.~\ref{sec:3}) both with interferometers and pulsar timing arrays (PTA).  

\noindent The formalism for the analysis of anisotropies of stochastic gravitational wave backgrounds (SGWBs) measurable
with ground and space-based detectors was introduced in \cite{Allen:1996gp,Cornish:2002bh}. Techniques developed for PTA can be found in \cite{Mingarelli:2013dsa,Hotinli:2019tpc}.  These studies are motivated by astrophysical phenomena:  anisotropies can for example be associated with groups of unresolved sources on  localised  regions of the sky, such as large cosmic structures. Similar  methods can also be applied in the context of our work, where anisotropies have a primordial origin and ${\cal Q}_{ij}$ is characterised by random matrix entries obeying Gaussian statistics. We shall adopt the   notation of the classic work   \cite{Allen:1997ad}. 
As first discussed in  \cite{Allen:1996gp}, the overlap function $\gamma_{12}$, associated with the cross-correlation of signals measured with a pair of ground-based detectors, receives  contributions due to  tensor anisotropies. In the vanishing frequency limit, and small antenna regime, we find a simple analytic expression for the  correction
 associated with the  quadrupolar anisotropy described by  Eq.~(\ref{q13}):
\begin{equation}
\label{18}
\gamma_{12}(f\to0)\,=\,2\,d_1^{ij} d_{2\,ij}-\frac87 {\cal Q}_{ij}\left(d_1^{\,im} \,d_{2\,m}^{j}+d_2^{\,im} \,d_{1\,m}^{j} \right)\,.
\end{equation}
Here $d_a^{ij}\,\equiv\,1/2 \,\left(\hat X_a^i \hat X_a^j -\hat Y_a^i \hat Y_a^j \right)$ denotes the  detector tensor, and ${\hat X}_a$, $\hat Y_a$  the interferometer arm directions. At high frequencies, the  contributions of the anisotropy to the overlap function are suppressed, and one recovers the results of   \cite{Allen:1997ad}. Anisotropies of  SGWBs  can then be detected and analyzed through their distinctive effects on a daily modulation of the signal, as
first proposed in   \cite{Allen:1996gp}.

\noindent
 One might wonder how to distinguish, when probing interferometer scales, primordial sources of quadrupolar tensor anisotropy from astrophysical ones. We stress that a bispectrum with a large component in the squeezed configuration can induce anisotropies in  the GW spectrum both at CMB and at interferometer scales. If the signal is sufficiently large to be measurable by  two independent probes, one may search for  common properties in the tensor quadrupolar harmonics, which may hint to a primordial origin for the anisotropies. We leave such investigations for future work.

\section{GW propagation and ultra-squeezed bispectrum}
\label{sec:3}

\noi We have seen in Sec.~\ref{sec2} how a long wavelength tensor mode can induce anisotropies in the  power spectrum.  At CMB scales the quadrupole serves as an indirect probe of squeezed non-Gaussianity, complementary to direct measurements of three-point correlations of temperature and polarization anisotropies.

 Is the same possible at small scales? Two recent works \cite{Bartolo:2018evs,Bartolo:2018rku} have shown that primordial tensor non-Gaussianity {\it cannot} be probed directly, i.e.  by measuring three (or higher, connected) point functions of tensor fluctuations at interferometer scales. Paraphrasing \cite{Bartolo:2018evs}, measurements of primordial tensor modes correlations at small scales involve angular integrations of contributions from signals produced by a large number of separate, independent, Hubble patches. In light of the central limit theorem, the statistics of tensor perturbations measured at interferometer scales will then be Gaussian. Even in the case of a set of detectors  built with the specific purpose of probing a large number of Hubble patches, one would not be able to  detect non-Gaussian correlations. Indeed, tensor non-Gaussianity at small scales is  suppressed due to Shapiro time-delay effects associated with the propagation of tensor modes at sub-horizon scales in the presence of matter. 
 Reference \cite{Bartolo:2018evs} suggests that observables sensitive to large correlations between short and long wavelength tensor modes, induced for example by an ultra-squeezed bispectrum,  may escape these conclusions. A concrete  realisation of such a possibility is precisely the quadrupolar anisotropy of the power spectrum discussed in Sec.~\ref{sec2}, which relied in part on previous works for the scalar \cite{Dimastrogiovanni:2014ina,Dimastrogiovanni:2015pla} and tensor
\cite{Ricciardone:2017kre,Dimastrogiovanni:2018uqy,Ozsoy:2019slf} cases. The quadrupolar asymmetry survives the aforementioned cancellation effects because it is induced by a \textit{super-horizon} tensor fluctuation. As we shall see in some detail, such mode
does not experience the  sub-horizon evolution responsible for the suppression of the non-Gaussian  signal.

 \subsection{Propagation at sub-horizon scales}

\noi We start by reviewing  how propagation affects short-wavelength modes \cite{Bartolo:2018rku}
 and then apply the same techniques to the case under scrutiny. Their momenta being centered at (ground or space-based) interferometers frequencies, short modes have entered the horizon during the radiation-dominated era. For $k>k_{\text{eq}}$ and $\eta<\eta_{\text{eq}}$, the mode-function reads 
\begin{equation}\label{1}
\gamma_{\textbf{k}}^{\text{RD}}(\eta)=j_{0}(k\eta) \gamma^{\text{prim}}_{\textbf{k}}
\,,
\end{equation}
where $j_{0}(k\eta)=\sin (k\eta)/(k\eta)$ is the spherical Bessel function and $\gamma_{\textbf{k}}^{\text{prim}}=\gamma_{k}^{\text{prim}}a_{\textbf{k}}+\gamma_{k}^{\text{prim}*}a_{-\textbf{k}}^{\dagger}$ is the primordial tensor perturbation. The initial conditions for $\gamma^{\text{prim}}$ are set by  inflation \footnote{The models that can be tested are those generating a squeezed bispectrum. A possibility is the model in \cite{Bordin:2018pca}, where  tensors are sourced linearly by extra (light) helicity-2 modes.  In the set-up of \cite{Dimastrogiovanni:2018uqy}  tensors are instead sourced quadratically.}.
Eq.~(\ref{1}) has been derived from the the tensor modes equation of motion
\begin{equation}
\gamma_{k}''+2\mathcal{H}\gamma_{k}'+k^2\gamma_{k}=0\,,
\end{equation}
by setting $\mathcal{H}=1/\eta$ for the radiation-dominated era. \\
During matter-domination, and to leading order in $|k\eta|$, one has \footnote{The formula below quantifies the Shapiro time-delay. It is obtained by modeling the presence of matter via interactions between $\gamma$ and $\zeta$ and working in the geometrical optic approximation: the scalar perturbation has a much longer wavelength than its tensor counterpart (see \cite{Bartolo:2018rku} for more details).},\cite{Bartolo:2018rku}:
\begin{equation}\label{2}
\gamma_{\textbf{k}}^{}(\eta)\simeq \frac{1}{(k\eta)^2}\left[\mathcal{C}(\textbf{k})\,e^{i\Gamma(\textbf{k},\eta)}+\mathcal{C}^{\dagger}(-\textbf{k})\,e^{-i\Gamma(-\textbf{k},\eta)}\right]\,,
\end{equation}
where $\mathcal{C}(\textbf{k})$ is a constant operator and 
\begin{eqnarray}
\Gamma(\textbf{k},\eta)&=&
  k \eta+2 k \int_{\eta_{\text{eq}}}^{\eta}d\eta'\Phi(\eta',(\eta'-\eta_{0})\hat{k})\,.
\\ \label{rela2}
&\equiv& k \eta+ Z(\textbf{k},\,\eta)
\end{eqnarray}
with $\Phi$ the Newtonian potential.  The above expressions underscore that inhomogeneities in the matter density at small scales -- encoded in
$\Phi$ --  affect the evolution of tensor modes. 
Matching Eqs.(\ref{1}) and (\ref{2}) at the time of matter-radiation equality, $\eta=\eta_{\text{eq}}$, implies 
\begin{eqnarray}\label{eqrr}
\gamma_{\textbf{k}}(\eta_{\text{eq}})&=&\frac{\cos(k\eta_{\text{eq}})}{(k\eta_{\text{eq}})^{2}}\left[\mathcal{C}(\textbf{k})+\mathcal{C}^{\dagger}(-\textbf{k})\right]\nonumber\\&&+\frac{\sin(k\eta_{\text{eq}})}{(k\eta_{\text{eq}})^{2}}\left[i\,\mathcal{C}(\textbf{k})-i\,\mathcal{C}^{\dagger}(-\textbf{k})\right]\nonumber\\&&=\frac{\sin(k\eta_{\text{eq}})}{(k\eta_{\text{eq}})}\gamma_{\textbf{k}}^{\text{prim}}\,.
\end{eqnarray}
From Eq.~(\ref{eqrr}) one obtains
\begin{equation}
\mathcal{C}(\textbf{k})=\frac{k \eta_{\text{eq}}}{2 i}\gamma_{\textbf{k}}^{\text{prim}}\,.
\end{equation}
The solution for $\eta>\eta_{\text{eq}}$ then becomes 
\begin{equation}\label{3}
\gamma_{\textbf{k}}(\eta)\,= \,\frac{\eta_{\text{eq}}}{k\eta^2}\left(\frac{e^{i \Gamma(\textbf{k},\eta)}-e^{-i \Gamma(-\textbf{k},\eta)}}{2\,i}\right) \gamma_{\textbf{k}}^{\text{prim}}
\,.
\end{equation}
Let us now compute the two-point correlation function:
\begin{eqnarray}
\langle \gamma^{\lambda_{1}}_{ij}(\textbf{k}_{1},
\eta) \gamma^{\lambda_{2}}_{ij}(\textbf{k}_{2},
\eta)\rangle&=&\frac{\eta_{\text{eq}}^{2}}{\eta^{4}k_{1}k_{2}}\frac{\epsilon_{ij}^{\lambda_{1}}(\hat{k}_{1})\epsilon_{ij}^{\lambda_{2}}(\hat{k}_{2})}{(-4)}\\&&\times \langle\gamma^{\text{prim}}_{\textbf{k}_{1}}\gamma^{\text{prim}}_{\textbf{k}_{2}}  \rangle\cdot\mathcal{E}(\textbf{k}_{1},\textbf{k}_{2},
\eta)\,,\nonumber
\end{eqnarray}
where
\begin{equation}\label{zZ}
\mathcal{E}\equiv \big\langle \left(e^{i \Gamma(\textbf{k}_{1},\eta)}-e^{-i \Gamma(-\textbf{k}_{1},\eta)}\right)\cdot\left(e^{i \Gamma(\textbf{k}_{2},\eta)}-e^{-i \Gamma(-\textbf{k}_{2},\eta)}\right) \big\rangle\,.
\end{equation}
Upon using 
\bea &\langle\gamma^{\text{prim}}_{\textbf{k}_{1}}\gamma^{\text{prim}}_{\textbf{k}_{2}}  \rangle=(2\pi)^3\delta^{\lambda_{1}\lambda_{2}}\delta^{(3)}(\textbf{k}_{1}+\textbf{k}_{2})P^{\lambda_{1}}(k_{1})\,,\nonumber \\
&\epsilon_{ij}^{\lambda_{1}}(\hat{k}_{1})\epsilon_{ij}^{\lambda_{1}}(-\hat{k}_{1})=1\; ,
\eea 
one finds
\begin{eqnarray}\label{5}
&&\langle \gamma^{\lambda_{1}}_{ij}(\textbf{k}_{1},
\eta) \gamma^{\lambda_{2}}_{ij}(\textbf{k}_{2},
\eta)\rangle=\nonumber\\
&&(2\pi)^3\delta^{\lambda_{1}\lambda_{2}}\delta^{(3)}(\textbf{k}_{1}+\textbf{k}_{2})P^{\lambda_{1}}(k_{1})\frac{\eta_{\text{eq}}^{2}}{(-4)\eta^{4}k_{1}^{2}} \,\mathcal{E}(\textbf{k}_{1},-\textbf{k}_{1},
\eta)\,.\nonumber\\
\end{eqnarray}
\noi Using Eq.~\eqref{rela2}, Eq.~(\ref{zZ}) becomes
\begin{eqnarray}\label{qQq}
\mathcal{E}&=&-2+e^{2ik_{1}\eta} \big\langle e^{iZ(\textbf{k}_{1},\eta)}e^{iZ(-\textbf{k}_{1},\eta)} \big\rangle\\&&+e^{-2ik_{1}\eta}\big\langle e^{-iZ(-\textbf{k}_{1},\eta)}e^{-iZ(\textbf{k}_{1},\eta)} \big\rangle\,.\nonumber
\end{eqnarray}
In Eq.~(\ref{qQq}), the expectation values can be computed using the relation \cite{Bartolo:2018rku}
\begin{equation}
\langle e^{\varphi_{1}}e^{\varphi_{2}}\rangle=e^{\frac{\langle\varphi_{1}^{2}\rangle}{2}+\frac{\langle\varphi_{2}^{2}\rangle}{2}+\langle \varphi_{1}\varphi_{2}\rangle}\,,
\end{equation}
which assumes Gaussian statistics for $\varphi$, yielding real exponentials multiplied by $ \cos(2k_{1}\eta)$, $\sin(2k_{1}\eta)$ functions. These terms drop out when performing the time average. One is  left with the contribution  
\begin{eqnarray}\label{unaffected}
&&\langle \gamma^{\lambda_{1}}_{ij}(\textbf{k}_{1},
\eta) \gamma^{\lambda_{2}}_{ij}(\textbf{k}_{2},
\eta)\rangle=\nonumber \\
&&(2\pi)^3\delta^{\lambda_{1}\lambda_{2}}\delta^{(3)}(\textbf{k}_{1}+\textbf{k}_{2}) \frac{\eta_{\text{eq}}^{2}}{2\eta^{4}k_{1}^{2}} P^{\lambda_{1}}(k_{1})\,.
\end{eqnarray}
\noindent We now extend these results and  consider the local power spectrum of gravitational waves evaluated in the presence of long-wavelength tensor perturbations, at a generic time $\eta$ during matter domination:
\begin{eqnarray}
\langle \gamma^{\lambda_{1}}_{ij}(\textbf{k}_{1},
\eta) \gamma^{\lambda_{2}}_{ij}(\textbf{k}_{2},
\eta)\rangle_{\gamma_{L}}&=&\frac{\eta_{\text{eq}}^{2}}{\eta^{4}k_{1}k_{2}}\frac{\epsilon_{ij}^{\lambda_{1}}(\hat{k}_{1})\epsilon_{ij}^{\lambda_{2}}(\hat{k}_{2})}{(-4)}\\&\times& \langle\gamma^{\text{prim}}_{\textbf{k}_{1}}\gamma^{\text{prim}}_{\textbf{k}_{2}}  \rangle_{\gamma_{L}}\cdot\mathcal{E}(\textbf{k}_{1},\textbf{k}_{2},
\eta)\,.\nonumber
\end{eqnarray}
Here Eq.~(\ref{3}) has been used and $\langle\gamma^{\text{prim}}_{\textbf{k}_{1}}\gamma^{\text{prim}}_{\textbf{k}_{2}}  \rangle_{\gamma_{L}}$ is given by Eq.~(\ref{q1}). The standard (isotropic) term in the first line of Eq.~(\ref{q1}) produces a contribution identical to the one in Eq.~(\ref{unaffected}). The term in Eq.~(\ref{q1}) proportional to the squeezed primordial bispectrum  becomes instead  
\begin{eqnarray}\label{771}
	&\langle \gamma^{\lambda_{1}}_{ij}(\textbf{k}_{1},
	\eta) \gamma^{\lambda_{2}}_{ij}(\textbf{k}_{2},
	\eta)\rangle_{\gamma_{L}}\supset
	-\frac{1}{4} \frac{\eta_{\text{eq}}^2}{\eta^{4} k_{1}k_{2}} 
	\epsilon_{ij}^{\lambda_{1}}(\hat{k}_{1})\epsilon_{ij}^{\lambda_{2}}(\hat{k}_{2}) \nonumber\\
	&\times
	\sum_{\lambda_{3}}\int_{|\vec{q}| < q_L} d^{3}q\,\delta^{(3)}(\textbf{k}_{1}+\textbf{k}_{2}+\textbf{q})\frac{\gamma^{*\lambda_{3}}_{\textbf{q}}\,B_{\gamma}^{\lambda_{1}\lambda_{2}\lambda_{3}}(k_{1},k_{2},q)}{P^{\lambda_{3}}_{\gamma}(q)} \nonumber\\
	&\times\mathcal{E}(\textbf{k}_{1},\textbf{k}_{2},
	\eta)\,.
\end{eqnarray}
Let us expand the expectation value in the last line of (\ref{771}) using Eq.(\ref{rela2})
\begin{eqnarray}\label{l22}
\mathcal{E}(\textbf{k}_{1},\textbf{k}_{2})&=&\big\langle e^{i\left(k_{1}+k_{2}\right)\eta}e^{i Z(\textbf{k}_{1},\eta_{})}e^{i Z(\textbf{k}_{2},\eta_{})}\nonumber\\&+&e^{-i\left(k_{1}+k_{2}\right)\eta}e^{-i Z(-\textbf{k}_{1},\eta_{})}e^{-i Z(-\textbf{k}_{2},\eta_{})}
\nonumber\\&+&e^{-i\left(k_{1}-k_{2}\right)\eta}e^{-i Z(-\textbf{k}_{1},\eta_{})}e^{i Z(\textbf{k}_{2},\eta_{})}\nonumber\\&+&e^{i\left(k_{1}-k_{2}\right)\eta}e^{i Z(\textbf{k}_{1},\eta_{})}e^{-i Z(-\textbf{k}_{2},\eta_{})}\big\rangle\,.
\end{eqnarray}
Similarly to what happens for the standard power spectrum, the contributions proportional to $e^{\pm i (k_{1}+k_{2})\eta}$ (first two lines of Eq.~\ref{l22}) average out because of the fast oscillation. Indeed, the period of oscillation, $\sim \pi/k_{1}$, is many orders of magnitude smaller than the integration interval. If we could set $\textbf{k}_{1}=-\textbf{k}_{2}$  the last two terms of (\ref{l22}) would become constant and constitute the only contributions left after averaging over time (as is the case for the standard power spectrum). This is precisely what happens in our context: from Eq.~(\ref{771}) we learn that the relation between the wavenumbers is $\textbf{k}_{1}=-\textbf{k}_{2}-\textbf{q}$, with $k_{1}\simeq k_{2}\gg q$. The difference $k_{1}-k_{2}$ is therefore of the order of the inverse cosmic time, $1/\eta_{0}$ ($q$ being at least horizon-size) and the exponentials $e^{\pm i(k_{1}-k_{2})\eta}$ can be treated as constants when performing the time average. Moreover, in the ultra-squeezed configuration ($\textbf{k}_{1}\simeq -\textbf{k}_{2}$) the arguments of the $Z$ terms in the last two lines in (\ref{l22}) are approximately equal and, from Eq.~(\ref{qQq}), one can thus set $\mathcal{E}(\textbf{k}_{1},\textbf{k}_{2})\simeq -2$. It follows that the quadrupolar anisotropy of the tensor spectrum, induced by the ultra-squeezed component of the tensor bispectrum, is not suppressed by
 propagation effects.

 \subsection{Averaging}

\noi As noted in \cite{Bartolo:2018evs}, the contributions of a primordial bispectrum to the three-point function of the detector time delay
  $\Delta\eta(\eta_{0})$, as measured along the interferometer arms,   vanishes as a result of rapidly oscillating phases $e^{i{\sum_{i}\pm k_{i}\eta_{0}}}$.  
This is not the case for the power spectrum: by enforcing $k_{1}=k_{2}$ the rapidly oscillating coefficient drops out \cite{Bartolo:2018evs}. Let us now include the contribution due to coupling with long-wavelength tensor fluctuations to the time delay two-point function. We find:
\begin{eqnarray}\label{eqnf}
\langle \Delta\eta(\eta_{0})\Delta\eta(\eta_{0})\rangle&\sim&\int d^{3}k_{1}d^{3}k_{2}\,e^{i\left(\textbf{x}_{1}\cdot\textbf{k}_{1}+\textbf{x}_{2}\cdot\textbf{k}_{2}\right)}e^{i\left(k_{1}-k_{2}\right)\eta_{0}}\nonumber\\&&\times\mathcal{M}\left(\hat{L}_{1}\cdot\hat{k}_{1},k_{1}\right)\mathcal{M}^{*}\left(-\hat{L}_{2}\cdot\hat{k}_{2},k_{2}\right)\nonumber\\&&\times\langle \gamma(\textbf{k}_{1},
\eta_{0})\gamma(\textbf{k}_{2},\eta_{0})\rangle_{\gamma_{L}}\,,
\end{eqnarray}
with ${\cal M}$ the detector transfer function, and $\hat L_i$ the interferometer arm direction.
The expectation value in the last line of Eq.~(\ref{eqnf}) includes, in addition to the diagonal contribution, also the off-diagonal term proportional to the primordial bispectrum in the ultra-squeezed limit. For the ultra-squeezed configuration (with the long-wavelength mode $q$ being at least horizon-sized), one has $|k_{1}-k_{2}|\simeq q$, with $q\leq \eta_{0}^{-1}$ , where $\eta_{0}$ is the cosmic time. As a result, $|k_{1}-k_{2}|\eta_{0}\simeq q\,\eta_{0}\leq 1$ and, similarly to the isotropic contribution, no suppression occurs. The  part controlled by the  squeezed bispectrum  selectively picks up the contribution of signals emitted from the same Hubble patch, without involving correlations from distinct Hubble regions.  \\

\noi We conclude that the quadrupole anisotropy computed in Sec.~\ref{sec2} propagates all the way to the observed tensor power spectrum, and hence the gravitational waves energy density.

\section{Conclusions}
\label{conclusions}
\noi The inflationary paradigm stands as one of the main pillars of modern cosmology. This position has been secured in light of its explanatory power on early universe dynamics and the agreement of inflationary predictions with observations. These successes notwithstanding, the current one is only a broad-brush picture of the inflationary mechanism with key questions still unanswered: what is the energy scale of inflation? What about its particle content? The observables that hold the most promise to access such information are the power spectrum and bispectrum of primordial correlation functions, both in the scalar and tensor sectors. The predictions corresponding to the two and three-point functions in the minimal inflationary scenario have long been known. Any observed deviation would therefore point directly to new physics.  

Currently available data place strong constraints on the inflationary scalar sector at CMB scales: the power spectrum  amplitude and scale dependence are known whilst non-Gaussianity is strongly constrained. The latter may still store key information on the mass, spin, and coupling of the theory field content. Many  more unkowns characterise the tensor sector with the predicted primordial signal still undetected and a tensor-to-scalar ratio correspondingly bounded to $r<0.06 \,(95\% {\rm CL})$ \cite{Akrami:2018odb}. The advent of ground and space-based laser interferometers makes it possible  to test for inflationary models with a non-standard scale dependence in the power spectrum and search for tell-tale signs of their particle content.

Recent studies \cite{Bartolo:2018evs,Bartolo:2018rku} have shown how  the signal from one key observable when it comes to probing extra fields, namely the primordial tensor bispectrum, is strongly suppressed at interferometers scales. Among other effects, propagation through structure de-correlates primordial modes of different wavelengths. Accessing the bispectrum    directly at interferometer (or e.g. PTA) scales, necessarily implies that all the modes have undergone (a long) sub-horizon evolution. In this work we  have put forward a complementary approach: looking for quadrupolar anisotropies in the tensor power spectrum as a probe of the primordial tensor bispectrum.  Despite not directly accessing the bispectrum, the quadrupole is nevertheless sensitive to modulations of an horizon-size tensor mode on the GW power spectrum. In this configuration one mode is insensitive to propagation  while the remaining two are very similar, thus avoiding an overall strong suppression. Quadrupolar tensor anisotropies can be probed at widely different frequency ranges and are therefore an efficient tool for testing inflationary models that support an ultra-squeezed component of the tensor bispectrum.

\section{Acknowledgements}
\label{acknowledgements}
\noi We are delighted to thank Valerie Domcke and Toni Riotto for comments on the manuscript, and David Wands for discussions on the subject. GT would also like to thank Nicola Bartolo, Ogan \"Ozsoy, Marco Peloso, Angelo Ricciardone, Toni Riotto for discussions. The work of MF is supported in part by the UK STFC grant ST/S000550/1. The work of GT is partially supported by STFC grant ST/P00055X/1.

\end{document}